\documentclass[11pt]{article}

\usepackage{graphicx}
\def\Journal#1#2#3#4{{#1} {\bf #2}, #3 (#4)}

\def\PRD{{\em Phys. Rev.} D}

\def\EPJ{{\em European Physical Journal} C}

\begin{document}
\title{Simulation of the SONGS Reactor Antineutrino Flux Using DRAGON}

\author{ Christopher L. Jones }


\maketitle
\begin{abstract}
For reactor antineutrino experiments, a thorough understanding of the fuel composition and isotopic evolution is of paramount importance for the extraction of $\theta_{13}$.  To accomplish these goals, we employ the deterministic lattice code DRAGON, and analyze the instantaneous antineutrino rate from the San Onofre Nuclear Generating Station (SONGS) Unit 2 reactor in California.  DRAGON's ability to predict the rate for two consecutive fuel cycles is examined.
\end{abstract}

\section{Motivation}
Recent experiments, such as the CHOOZ experiment [1], have been able to set lower bounds on the determination of $\theta_{13}$:  

\[ \sin^2\theta_{13} < 0.19  \]

The experiment has also indicated that reactor-based uncertainties have become the dominant systematic error in extracting $\theta_{13}$.  In a breakdown of contributions to the overall systematic errors [2], it is clearly seen that the error on the antineutrino flux outweighs the others.  Thus, if one is to decrease the $\theta_{13}$ lower bound, one must contend with and accurately model this flux; this requires detailed simulation of the reactor internals.  Prediction of the flux is our final goal; to be confident in our predictions, however, we must benchmark our code against existing real data.  To expedite the benchmarking process, instead of using flux data, we use measurements of the antineutrino rate from a reactor since we were in ready possession of this data.  Simulation of the rate and flux both require our code to predict a set of key quantities: the \textit{fission rates} of the four radioisotopes that contribute $> 99$\% of the fissions in a nuclear reactor.  If our rate prediction is adequate, then we can immediately apply our results to a flux prediction when that data is available.  Thus, for the rest of this report, we will focus only on the rate prediction, particularly on the determination of the fission rates.  First, we give a brief overview of the reactor and detector from which we have obtained our rate data.

\section{SONGS: Reactor and Detector}\label{subsec:songs}
\subsection{Reactor}
SONGS Unit 2 is a PWR (pressurized water reactor) in California.  It is operated by Southern California Electric (SCE) and has a thermal power output of 3.46 GW.  Our colleagues at Lawrence Livermore National Laboratory (LLNL) were able to obtain very detailed inputs and outputs of their simulation package that were then employed as inputs to our own reactor simulation.  These SCE values were not accompanied by uncertainties, so it will be up to us in the future to assign and / or estimate the associated uncertainties to these variables.

The reactor core is divided into 217 subunits called \textit{fuel assemblies}.  Each of these fuel assemblies is a bundle of fuel rods (explaining a common alternate name, the \textit{fuel bundle}).  The fuel assembly is a ``degree of freedom" for the reactor core; after a fuel-burning cycle, the assembly as a whole can be removed, refreshed, and reloaded.  This subdivision is also convenient since our DRAGON code also specializes in assembly simulation. 
This reactor is labeled ``16 x 16" because it is comprised of a 16-by-16 grid of fuel rods (236) and instrumentation rods (20).  These fuel rods contain ``fresh" fuel (only uranium) as well as ``burned" mixtures of fuel (uranium with plutonium and other heavy metals).  Detailed information about each assembly, from their temperature during fission to their composition of the fuel cladding, was provided for the simulation.

\subsection{Detector}
The SONGS detector, commissioned and operated by LLNL and Sandia Laboratory, contains approximately 0.64 metric ton of gadolinium-doped scintillator, located  24.5 meters from the SONGS reactor core.  The detection mechanism uses the tried-and-true inverse $\beta$ decay process, whose products produce a delayed coincidence signal between the annihilating positron and the neutron capture; the very high thermal capture cross section of gadolinium enhances this signal.  The overall detection efficiency is 10.7\%, which is itself known to approximately 10\%.
With this information, we can now simulate the SONGS reactor, the topic to which we turn next.

\section{The DRAGON Code}

	DRAGON is an open-source simulation package that allows one to study the behavior of neutrons in a nuclear reactor.  It allows one to determine the isotopic concentrations of radionuclides during the burnup cycle, as well as to perform isotopic depletions.  The open source nature of this code is attractive not only because of cost factors, but also due to the relative ease of its modification.  Indeed, \textbf{our DRAGON simulation package has been modified to output the fission rates of the four most important isotopes in fissions, $^{235}U, ^{238}U, ^{239}Pu$, and $^{241}Pu$}.  Obtaining these fission rates as a function of time is the primary use of DRAGON.
	
Since DRAGON is a deterministic code, not a Monte Carlo code (such as MCNP), there is no error associated with its calculations and we must estimate this ourselves.  DRAGON is also known as a ``2D" code, since it only simulates a 2D cross section of the fuel assembly.  A full 3D solution of the multidimensional neutron diffusion equation is very expensive deterministically, so DRAGON provides the option of interfacing its 2D solution with its companion code, DONJON, to do a 3D calculation.  However, as a first simplification, we have merely scaled the fission rates from DRAGON by the height of the reactor ($H$ = 381 cm).  It has been checked that this procedure is appropriate unit-wise.  While it makes the assumption that the fission rates--and thus the flux--will be uniform along the axial direction when they are not, we note that the CHOOZ reactor simulations also made this uniform axial flux assumption [3], and ours can be relaxed in the future with DONJON.
With the fission rates, we can calculate the antineutrino rate.

\section{The SONGS Detected Antineutrino Rate}

\begin{figure}
\begin{center}
\includegraphics[scale=0.6]{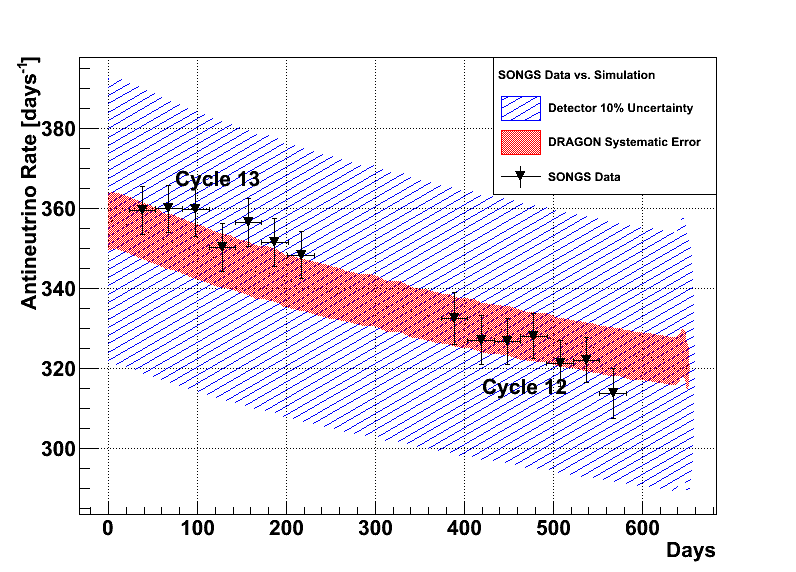}
\caption{SONGS Rate Prediction With DRAGON}
\end{center}
\end{figure}

DRAGON's prediction of the rate detected at SONGS is given by 

\[  \frac{dN_\nu}{dt} = \frac{\epsilon N_p}{4\pi D^2} \sum_i f_i \int_{1.806}^{\infty} \,dE_\nu \sigma(E_\nu) S_i(E_\nu)  \]

Here, $i$ sums over the aforementioned isotopes, $f_i$ are fission rates, $\sigma(E_\nu)$ is the inverse $\beta$ decay cross section, $D$ is the distance to the reactor, $N_p$ is the number of target protons (= 4.35 $\times 10^{28}$) in the detector, and $\epsilon$ is the detection efficiency.  The functions $S_i$ are the parameterizations of the \textit{Schreckenbach spectra}, which provide the number of antineutrinos produced per fission, per radionuclide.  This parameterization is provided by Petr Vogel [4].

\subsection*{Discussion}
The work on this rate calculation has progressed since the time of the conference and the plot shown here differs greatly from the one shown there.   The normalization concerns mentioned during the author's talk have been corrected.  Also, this time two different fuel cycles are shown.  The SONGS data were taken at the latter half of Cycle 12; after the reactor was shut down and refueled, data were taken for the first half of the subsequent Cycle 13.  These cycles had different fuel loadings and thus different rate predictions.  

The plot here shows the prediction for Cycle 12 with a 2\% error band (red band); 2\% was chosen as a broad estimate since American reactors must be certain that their power output is within 2\%.  Due to the 10\% uncertainty in the detector efficiency (blue band), it is difficult to estimate DRAGON's effectiveness at predicting the rate.  However, there are other comparisons that we can make; one of these is to ascertain if DRAGON predicts the same isotopic concentrations at the end of its fuel cycle as predicted by SCE itself.  Work of this nature is underway.  Also, even if absolute normalization is lacking, information about the slope will suffice for some rudimentary nonproliferation efforts.

\section*{Acknowledgments}
The author would to thank parties at both MIT (Professor Janet Conrad and Dr. Lindley Winslow) and at LLNL (Drs. Adam Bernstein, Gregory Keefer, and Nathaniel Bowden) for their support and frequent discussions.  The author is especially grateful for the generous assistance of Professor Guy Marleau, a principal author of the DRAGON code, for support in the code modification.  The author  is supported by  NSF grant PHY-0847843.


\begin{thebibliography}{99}

\bibitem{ch}M. Apollonio {\it et.al.}, \Journal{\EPJ}{27}{331}{2003}.

\bibitem{zl} Z. Djurcic {\it et.al.}, arXiv:hep-ex/0808.0747v1 (2008).

\bibitem{dn}D. Nicolo, {\em Search for neutrino oscillations in a long baseline experiment at the CHOOZ nuclear reactors}, (Edizioni della Normale, 2007).

\bibitem{vg}P. Vogel, J. Engel, \Journal{\PRD}{39}{11}{1989}

\end{thebibliography}
\end{document}